\documentclass{aa}
\usepackage{graphicx}
\usepackage{amsfonts}
\begin{document}

\title{Buoyant magnetic flux ropes in a magnetized stellar envelope}

\subtitle{Idealized numerical 2.5-D MHD simulations}

\author{S.B.F. Dorch\inst{1,2}}

\offprints{S.B.F. Dorch, \email{dorch@astro.ku.dk}}

\institute{ Copenhagen University Library,
N{\o}rre all{\'e} 49, DK-2200 Copenhagen N, Denmark \and The Niels
Bohr Institute, Copenhagen University, Juliane Maries Vej 30,
DK-2100 Copenhagen {\O}, Denmark}

\date{Received date / Accepted date}

\authorrunning{Dorch}

\abstract
   {The context of this paper is buoyant toroidal magnetic flux ropes, which is a part of flux tube dynamo
   theory and the framework of solar-like magnetic activity.}
   {The aim is to investigate how twisted magnetic flux ropes interact with a simple magnetized
   stellar model envelope---a magnetic ``convection zone''---especially to examine how the twisted magnetic field
   component of a flux rope interacts with a poloidal magnetic field in the convection zone.}
   {Both the flux ropes and the atmosphere are modelled as idealized 2.5-dimensional concepts
   using high resolution numerical magneto-hydrodynamic (MHD) simulations.}
   {It is illustrated that twisted toroidal magnetic flux ropes can interact with a poloidal magnetic field in
   the atmosphere to cause a change in both the buoyant rise dynamics and the flux rope's geometrical shape.
   The details of these changes depend primarily on the polarity and strength of
   the atmospheric field relative to the field strength of the flux rope.
   It is suggested that the effects could be verified observationally.}
   {}


\maketitle

\section{Introduction}

Buoyant magnetic flux tubes are an essential part of the framework
of current theories of dynamo action in both the Sun and
solar-like stars: it is widely believed that flux tubes are formed
by some combination of rotational shear and turbulent
convection near the bottom of the convection zones (CZ) of these
stars. When sufficiently buoyant the magnetic flux rise in the
form of tubular (toroidal) $\Omega$-shaped loops of magnetic field
lines. Rising under the influence of rotational forces, the loops
finally emerge as slightly asymmetric and tilted bipolar magnetic
regions at the surface, see e.g.\ \cite{Fan+ea94} and
\cite{Caligari+ea95}. During the last decade, it has been
established that if they do exist, these flux tubes must in fact
be flux {\em ropes} of intertwined or twisted field lines;
otherwise they would quickly be disrupted by a magnetic
``mushrooming'' instability, see e.g.\ \cite{Emonet+Moreno1998}
and \cite{Dorch+Nordlund1998}. It has been shown that the
instability is inhibited if the degree of systematic twist is
sufficiently high, corresponding to a critical value of the field
line pitch angle $\Psi_c$ approximately determined by equality
between the energy density of the twisted field component and the
ram pressure due to the rise of the flux rope e.g.\
\cite{Moreno+Emonet1996}.

A fundamental ingredient in all theories of solar-like dynamos is
a process where a relatively weak poloidal magnetic field is
turned into a toroidal field by differential rotation: it is this
toroidal field that eventually gives rise to the flux ropes,
which at the end of their existence return a poloidal field
component to the CZ, thereby completing the magnetic activity
cycle. This dynamo process is gradual and different cycles
overlap.

Buoyant magnetic flux ropes have been extensively study
through numerical simulations, cf.\ the reviews in
\cite{Dorch2002} and \cite{Fan2004}, and the most recent high
resolution study in two dimensions by \cite{Cheung+ea2006}.
However, so far studies of the interaction of flux ropes with
poloidal fields have solely dealt with flux ropes that emerge into
the solar corona, e.g.\ \cite{Archontis+ea04}, i.e.\ not with a
magnetic CZ.

The question addressed in this paper is how the polarity and
strength of the field in a magnetized convection zone may affect
the rise and evolution of twisted flux ropes. When the
predominantly toroidal buoyant flux ropes rise through the CZ,
they encounter a poloidal magnetic field that has a component
perpendicular to the ropes' axes. However, since the flux ropes
are twisted, the transversal field components may be either
parallel or anti-parallel to the average magnetic field in the CZ,
i.e.\ the field in the two-dimensional plane perpendicular their
toroidal axis.

Several questions relevant to the theory of flux tube dynamos then
emerge: e.g.\ is the rise faster or slower and do more or less
magnetic flux reach the surface in the presence of a magnetized
convection zone; how does a magnetized convection zone affect the
required amount of twist. In this paper I attempt to shed light on
these questions by presenting results from idealized numerical
2.5-dimensional MHD simulations of the cross-sections of
buoyant flux ropes that interact with a horizontal magnetic layer
while they rise.

\section{Model}
\label{model.sec}

The full compressible MHD-equations are solved using the
stagger-code by Galsgaard and others, cf.\
\cite{Galsgaard+Nordlund1997}:
\begin{eqnarray}
{\frac{\partial {\bf \rho}} {\partial t}} & = & -\nabla\cdot\rho{\bf u}, \label{mass.eq}\\
{\frac{\partial {({\rho}{\bf u})}} {\partial t}}& = & -\nabla P
    + {\bf j}\times{\bf B} - \nabla\cdot(\rho{\bf u}{\bf u})
    + \nu \nabla^2 \rho {\bf u} \label{motion.eq},\\
{\frac{\partial e}  {\partial t}} & = & - \nabla\cdot(e{\bf u})
    - P\nabla\cdot{\bf u} + Q_{\rm v} + Q_{\rm J},
    \\ \label{energy.eq}
\frac{\partial {\bf B}}{\partial t} & =  &\nabla \times ( {\bf u}
\times {\bf B})
    + \eta \nabla^2 {\bf B}, \label{induction.eq}
\end{eqnarray}
Here $\rho$ is the fluid density, ${\bf u}$ is the velocity, $P$
the gas pressure, ${\bf j}$ is the electric current density,
${\bf B}$ is the magnetic field density and $e$ is the internal
energy. In Eq.\ (\ref{motion.eq}) ${\bf j}\times{\bf B}$ is the
Lorentz force and $Q_{{\rm v}}$ and $Q_{{\rm J}}$ are the viscous
and Joule dissipation respectively.

The equations are solved numerically on a staggered mesh using
derivatives and interpolations that are of 6th and 5th order
respectively in a numerical scheme that conserves $\nabla\cdot{\bf
B}=0$ exactly. The time stepping is implemented by a third order
predictor-corrector method. Numerical solutions are obtained on a
high-resolution two-dimensional Cartesian grid of $2048^{2}$
points. The initial set-ups of the models are twofold,
consisting of a snapshot of a stratified and magnetized adiabatic
CZ-model, and of an idealized twisted magnetic flux rope.

Initially the entropy in the interior of the ropes are set
equal to that in the external medium leading to a buoyancy equal
to $1/\gamma\beta$ (with $\gamma = 5/3$). The coordinate system is
chosen such that $x$ is the vertical coordinate, $y$ the
horizontal coordinate and $z$ is the coordinate along the axis
which is parallel to the rope.

The flux ropes move in a polytropic atmosphere $P=P_0
(\rho/\rho_0)^{\gamma}$, where $P_0$ and $\rho_0$ are the
quantities at the initial position of the flux rope corresponding
to a pressure scale height of ${\rm H}_{\rm P0} = 59$ Mm. The top
pressure scale height is one tenth of the bottom scale height
${\rm H}_{\rm P0}$. The upper boundary is well below the position
of an imagined photosphere. Horizontally the boundaries are
periodic and the physical size of the computational box is 136 Mm
in both the vertical and horizontal dimensions corresponding to a
height of 2.3 ${\rm H}_{\rm P0}$. The initial position of the flux
ropes are set to $x = 0.27~ {\rm H}_{\rm P0}$.

The initial twist of the flux ropes are given by
\begin{eqnarray}
 {\rm B}_z      & = & {\rm B}_{\rm 0} e^{-(r/ R)^2}\\
 {\rm B}_{\phi} & = & \alpha (r / R)^3~ {\rm B}_z e^{-(r/ R)^2}, \label{initial.eq}
\end{eqnarray}
where ${\rm B}_z$ is the parallel (axial) and ${\rm B}_{\phi}$ the
transversal component of the magnetic field with respect to the
ropes' main axes. ${\rm B}_{\rm 0}$ is the amplitude of the field,
$R$ the radius, and $\alpha$ is a field line pitch parameter. The
critical pitch of a twisted flux rope needed to prevent the
disrupting instability depends on the ratio of the rope's radius
to the local pressure scale height: for thin flux ropes this ratio
is small and in the case of the thin flux ropes studied by
\cite{Emonet+Moreno1998}, the critical pitch angle was $10^{\rm
o}$. In the following I set the initial radius to $R_0 = 0.14~
{\rm H}_{\rm P0}$, i.e.\ the tubes are non-thin yielding a
critical angle of $\Phi_c \approx 25^{\rm o}$ according to
the expression by \cite{Moreno+Emonet1996}. To be on the safe
side, $\alpha = 1.75$ is chosen, so that the initial maximum
field line pitch angle is $35.6^{\rm o}$, occurring at the border
of the rope at $r = \sqrt{3/2}~ R$ for the topology given in Eq.\
(\ref{initial.eq}). Incidentally this is low enough that a
three-dimensional flux rope would not be kink unstable, since to
achieve a substantial growth rate this requires a pitch exceeding
$45^{\rm o}$, e.g.\ \cite{Galsgaard+Nordlund1997}.

\begin{table}[!htb]
\caption[]{\small List of numerical simulations and their basic
parameters: model name, ratio of CZ field to the rope's twisted
field component $\epsilon$, sign of the latter ratio, and ratio of
the rope's axial field to the CZ field strength $\chi$ (only
the value of $B_{\rm CZ}$ vary). The horizontal dividing line
indicates the approximate position of the lower solar limit, i.e.\
$\chi_\odot \sim 10$.} \label{tab1} \centering
\begin{tabular}{l c c c}
\hline\hline
Model & $\epsilon$ & $||$ & $\chi$ \\
\hline
0  & 0   & N/A & N/A \\
1A & 0.044 & -   & 90  \\
1B & 0.044 & +   & 90  \\
2A & 0.22 & -   & 18  \\
2B & 0.22 & +   & 18  \\
\hline
3A & 0.44   & -   & 9   \\
3B & 0.44   & +   & 9   \\
4A & 2.2  & -   & 1.8 \\
4B & 2.2  & +   & 1.8 \\
\hline
\end{tabular}
\end{table}

With a grid of 2048 points in each dimension, the flux ropes are
resolved by more than 200 grid points while the resolution of the
pressure scale height ${\rm H}_{\rm P0}$ is closer to 1000 grid
points. From a computational point of view one has to set the
ropes' plasma $\beta$'s lower than the solar values
($10^4$---$10^7$) to reduce the computational time scale to a
reasonable value: $\beta_0 = 107$ is chosen for the initial
strength of the buoyancy on the ropes' axes.

The magnetic layer is implemented by adding a magnetic field
of $B_y = B_{\rm CZ}$ to the hydrostatic equilibrium at heights $x
\geq 0.9~ {\rm H}_{\rm P0}$ above the bottom of the computational
box. Numerically this results in a steep gradient at that
height (see the discussion on diffusion in Section 4).

For convenience I furthermore define $\chi \equiv  B_{\rm
rope}/B_{\rm CZ}$ for the ratio of the ropes' axial fields to the
CZ field strength, and $\epsilon \equiv B_{\rm CZ}/B_\phi$ for the
ratio of CZ field to a characteristic value of the ropes'
twisted field components, e.g.\ its maximum value. In the
simulations presented here both $\epsilon$ and $\chi$ are varied
by a factor of 50 between the weakest and strongest CZ field
strength. For all models $\epsilon \chi$ has the same value
$\approx 4$. An estimated value of $\chi$ for the Sun is given by
the ratio of the canonical toroidal field strength of 100 kG, to a
poloidal field of at most equipartition $B_{\rm eq} \sim 10$ kG,
i.e.\ $\chi_\odot \geq 10$.

It is trivial to calculate a simple limit for the strength of the
envelope field $B_{\rm CZ}$ relative to the strength of rope by
assuming a force balance between the buoyancy of the flux rope and
the oppositely directed tension caused by a dent of magnitude $R$
in field lines in the CZ field layer: with an initial buoyancy of
$1/\gamma\beta$ the order of magnitude of $\chi$ should
exceed
\begin{equation}
 \chi_{\rm c} \equiv \sqrt{ \gamma \frac{2~ {\rm H}_{\rm P0}}{{\rm
 R}}}, \label{chi_c.eq}
\end{equation}
if buoyancy is to overcome the magnetic tension from the field
lines in the layer. The size of the ropes in this model results in
$\chi_0 \sim 5$. For the Sun, the approximate lower bound fulfills
$\chi_\odot > \chi_{\rm c}$ and hence solar flux ropes are
expected to be able to rise through the poloidal field layer as
assumed by the flux tube dynamo.

\begin{figure*}[!htb]
\centering
 \includegraphics[width=17cm]{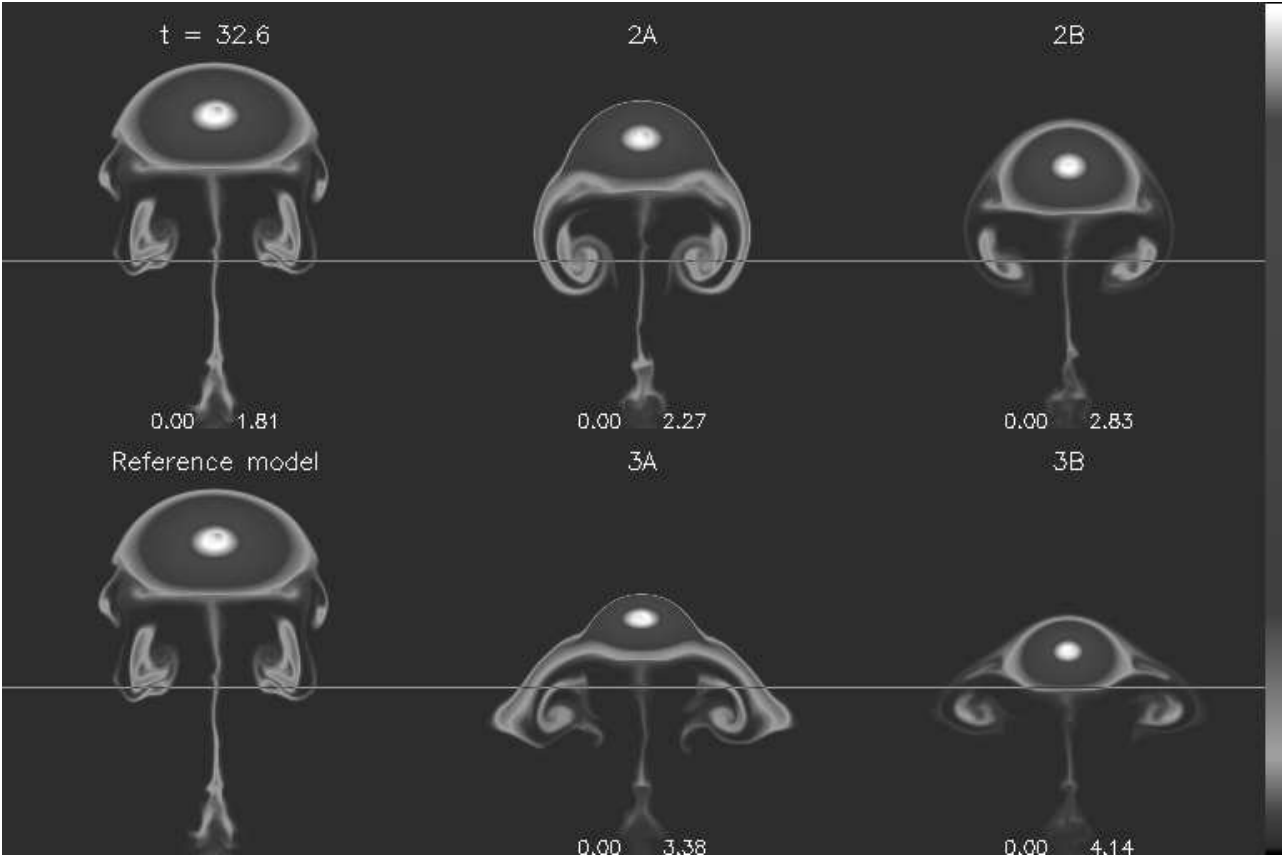}
\caption[]{Grey-scale panel showing snapshots of $B_z$ at $t \sim
32.6$ time units for models with moderately magnetized envelopes.
Minimum (zero) and maximum magnetic field strengths are stated for
every snapshot (in units of 354 kG). Vertical dimension
corresponds to $x$ in the simulations. Shown are model 0 (shown
twice), and in the upper row models 2A and 2B (initial $B_{\rm
CZ}/B_\phi = 18$) and in the lower row models 3A and 3B (initial
$B_{\rm CZ}/B_\phi = 9$). Also shown are horizontal lines
indicating the initial position of the magnetized layer.}
\label{fig3}
\end{figure*}

\section{Results }
\label{results.sec}

Several models were run, varying parameters such as numerical
resolution and $B_{\rm CZ}$, but not the ropes' initial field
strengths. Table \ref{tab1} lists the models that are discussed
in this paper. Except from models 4A and 4B all models have $\chi
> \chi_{\rm c}$.

The nine models consist of eight models of flux ropes rising into
a magnetized layer corresponding to a poloidal field in a CZ and a
single reference model where no CZ field was present model 0.

\subsection{Flux conservation and numerical dissipation}

With a weak or absent envelope magnetic field, the total magnetic
flux through the plane perpendicular to a rope's axis
$\Phi_{\rm z}$ remains constant: it only decreases by 0.1 promille
during the simulation runs and hence there is virtually no flux
loss due to (unwanted) numerical dissipation. The total magnetic
flux through the horizontal boundaries is another matter: the
total flux $\Phi_{\rm y}$ through the horizontal boundary
decreases constantly by typically 0.5 promille per time unit,
leading to a loss of 1--3\% during an entire run. This loss could
in theory be due to magnetic field being pushed through the upper
boundary. However, the upper boundary remains closed to the
passage of plasma and as will be discussed subsequently, the flux
loss results primarily from natural physical diffusion of the CZ
magnetic field layer.

\subsection{Buoyant rise of the ropes}

The single simulation (model 0, lacking a magnetized
stratification) is completely identical to that presented in the
review by \cite{Dorch2002} and elsewhere (albeit here in much
higher numerical resolution): as the rope rises and expands it
approaches a ``terminal rise phase'' where it rises with a speed
approaching a terminal velocity $v_t$ (determined by the force
balance between buoyancy and drag). In this phase the rope e.g.\
oscillates due to its differential buoyancy and rise more or less
in the same way as an adiabatically rising, non-stretching flux
tube obeying an internal polytropic equation of state, see
\cite{Dorch+Nordlund1998} and \cite{Dorch2002}.

For all the models in Tab.\ \ref{tab1}, the initial behavior of
the ropes is unaffected by the presence of the overlying
magnetized layer. When a rope begins to rise due to its buoyancy
however, it eventually approaches the layer: while the plasma in
front of the rope moves away to ``let the rope through", part of
the mass---in front of the rope (because of mass continuity, Eq.\
\ref{mass.eq})---moves upwards along with the rope and pushes on
the horizontal magnetic field lines, thereby causing an indent of
compressed field in the layer. At some point the tension in the
dented field lines becomes sufficient to withstand the compression
by the upward flow, and the forefront of the rope and the nearly
horizontal field lines begin to approach each other faster. In the
models, this happens at a time of $t \sim 50~ \tau_{\rm g}$, with
$\tau_{\rm g} = \sqrt{{\rm H}_{\rm P0}/g}$.

The most pronounced effect is clearly the case where the CZ field
is the strongest, corresponding to $\epsilon = 2.2$ in the
simulations model 4A and model 4B that have an envelope field of
the same order of magnitude as the initial strength of the flux
rope and hence $\chi < \chi_{\rm c}$. The rise of these ropes is
quickly halted and they never get past the threshold to the
magnetized envelope: they remain below the layer while performing
damped oscillations eventually becoming a perpendicular
(toroidal) magnetic layer, similarly to what happens to weak flux
tubes in photospherical simulations, e.g.\ \cite{Magara2001} and
\cite{Archontis+ea04}.

The simulation sets numbered 3A+3B, 2A+2B and 1A+1B respectively,
are more interesting in the solar context. These simulations
correspond to situations where the rope's axial strength is
approximately 10, 20 and 90 times that in the magnetized envelope:
this covers an order of magnitude in the ratio of the envelope's
field to the rope's twisted field component, $\epsilon$. Figure
\ref{fig3} compares $B_z$ for the two first of these simulation sets
to the reference model 0 at a late time ($t \sim
95~\tau_{\rm g}$): the effects of the presence of the poloidal layer
are clear, and these will be discussed in the following.

\begin{figure*}[!htb]
\centering
 \includegraphics[width=8.5cm]{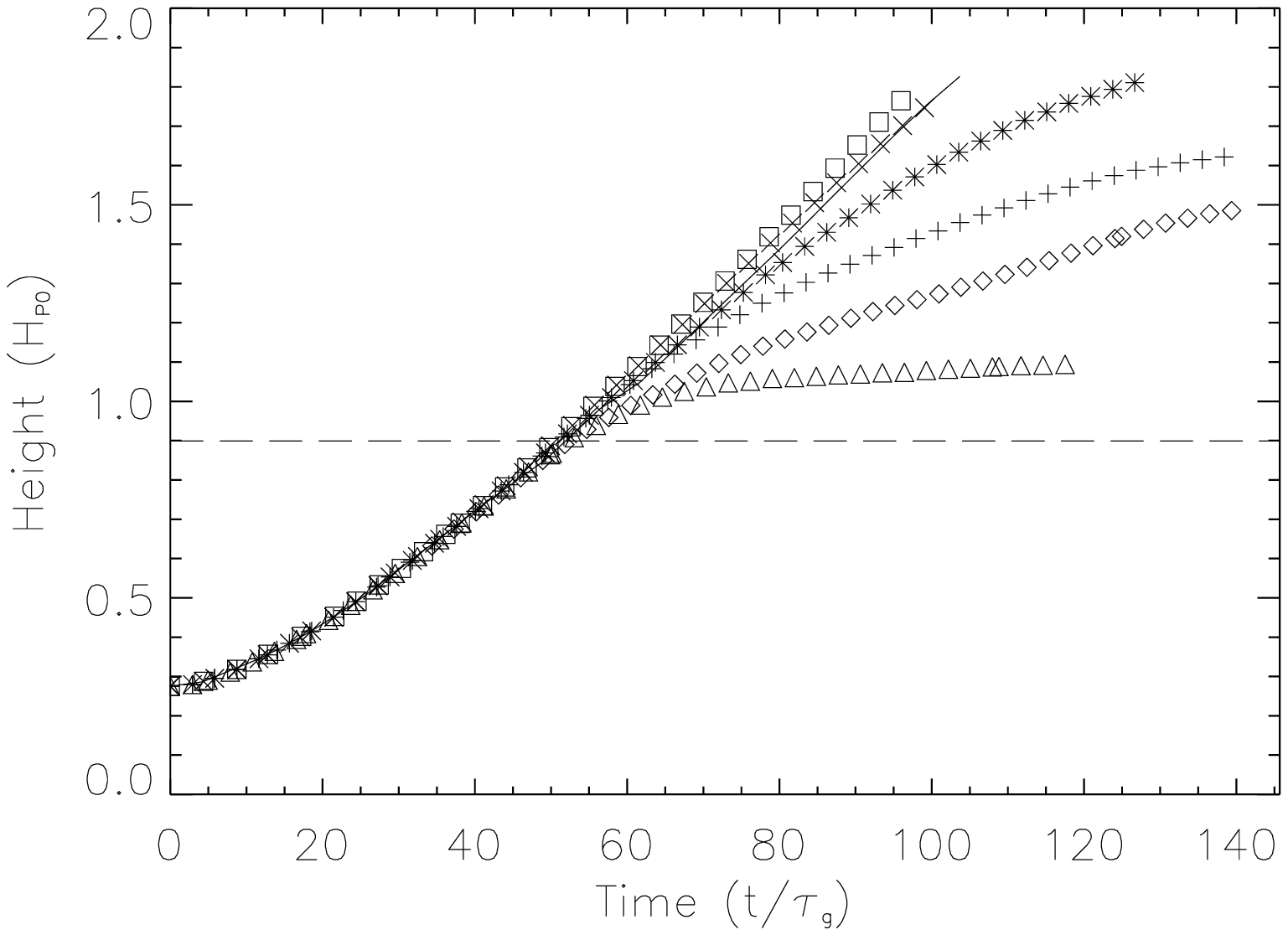}
 \includegraphics[width=8.5cm]{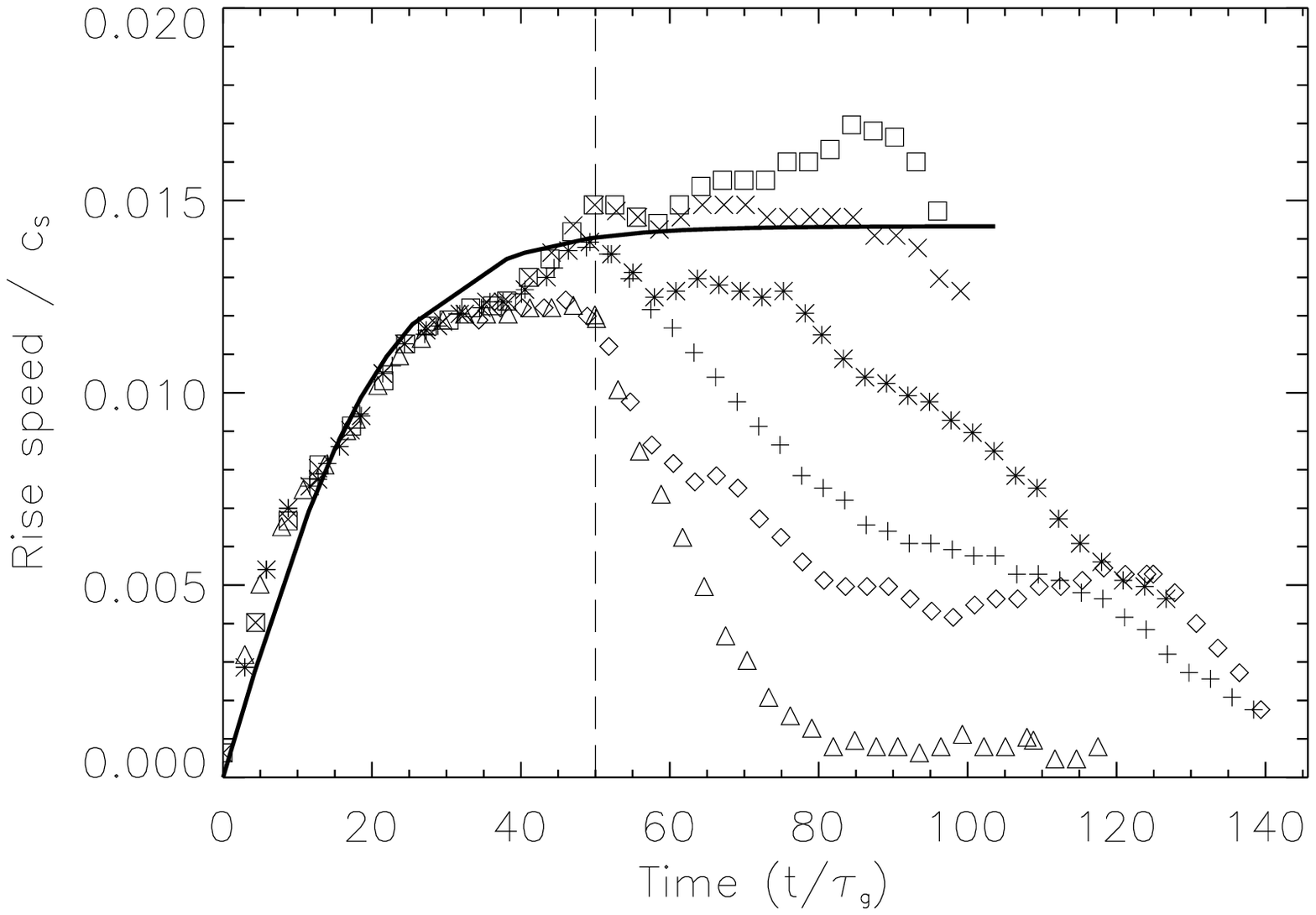}
\caption[]{Left: the height of the flux ropes (in terms of $H_{\rm
P0}$) as a function of time for six models indicated by symbols.
In order of increasing strength of the magnetic layer the models
are: 1A (squares), 1B (crosses), 2A (stars),
2B (plus), 3A (diamonds) and 3B (triangles).
Also shown is reference model 0 (solid line). The vertical
dashed line indicates the initial lower boundary of the magnetized
layer in the CZ. Right: speed of the flux ropes as a function of
time (in units of the sound speed at the initial position of the
ropes). Here the thick solid line illustrates the theoretically
expected asymptotic behavior of the rise speed for an un-magnetized
CZ. } \label{fig2}
\end{figure*}

Figure \ref{fig2} (left) shows the height of the flux ropes as a
function of time: in the case of $\epsilon = 0.22$ and $\epsilon =
0.44$ the CZ field lowers the speed of rise, compared to the
reference model. Neither of the two models with $\epsilon = 0.44$
reach very far before being halted within the magnetized layer,
most easily seen in case of model 3B in Fig.\ \ref{fig2} that has
an envelope-parallel field, where the flux rope is trapped at a
height of $\sim 1.1~ H_{\rm P0}$, just in side the magnetic layer.

Figure \ref{fig2} (right) shows the same trend in terms of the
rise speed: the expected evolution for a flux tube rising in a
field-free environment is an asymptotical approach to a terminal
rise speed. However, with a magnetic CZ corresponding to the high
$\epsilon$ values, after reaching a maximum, the rise speed
plummets to zero, when the ropes encounter the magnetized layer
at time $t \sim 50~ \tau_{\rm g}$. An exception may be model 2A
that has an anti-parallel twist field: at the end of the
simulation this rope still had a substantial non-zero speed,
although declining.

On the one hand, as one may naively expect in the four situations
just discussed, the parallel-field layers lower the speed more
than anti-parallel layers does. This is because of the
impossibility of reconnection between the two magnetic flux
systems in this case, that would otherwise lower field line
tension by removing transversal field from the region above the
apex of the flux ropes. On the other hand, consider the cases with
$\epsilon = 0.044$ and $\chi = 90$: these ropes actually rise
marginally faster than the CZ field-free reference model, while
again the simulation with an anti-parallel twist rises the fastest
of the two (see Fig.\ \ref{fig2}).

\begin{figure*}[!htb]
\centering
 \includegraphics[width=8.5cm]{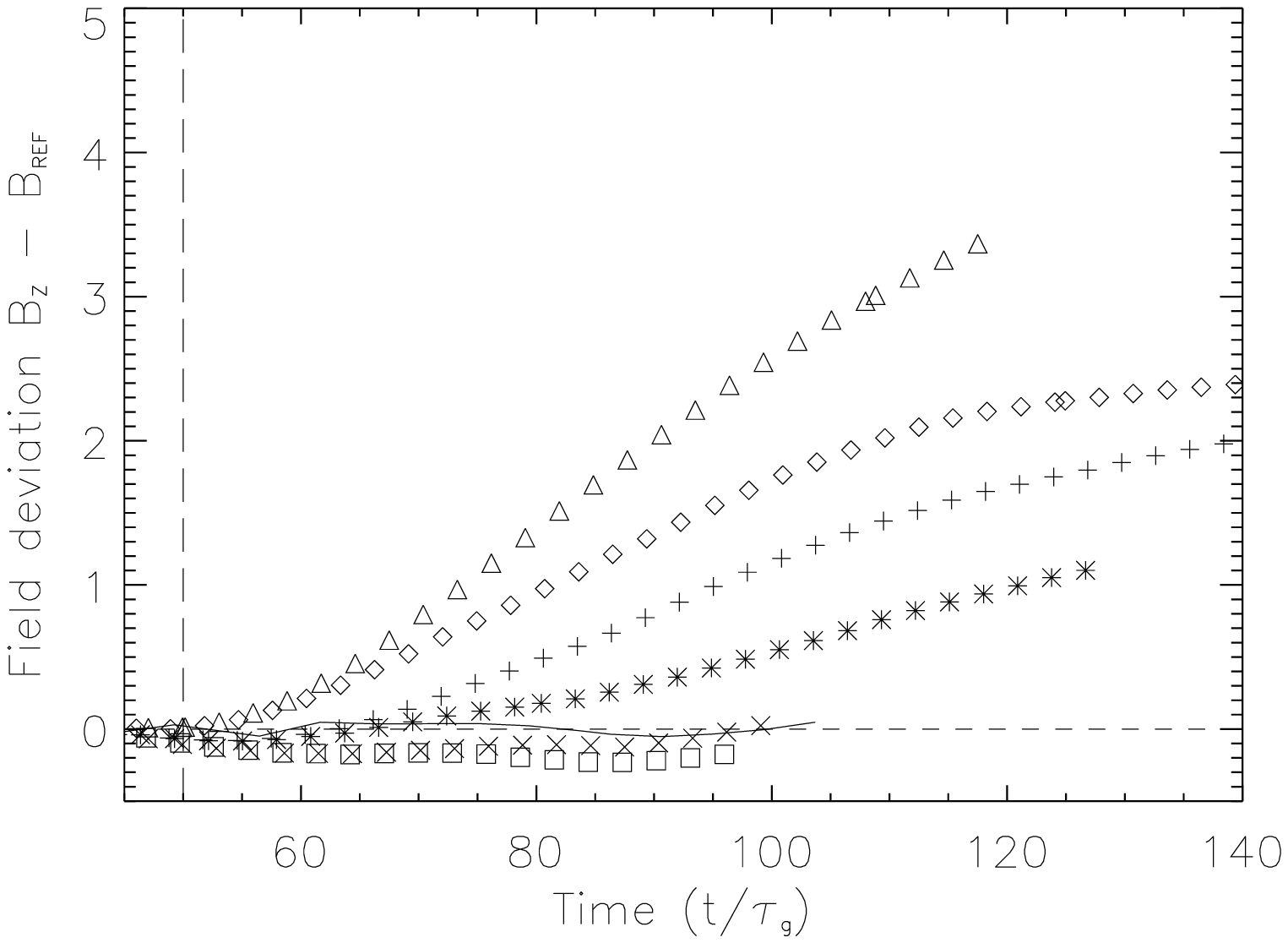}
  \includegraphics[width=8.5cm]{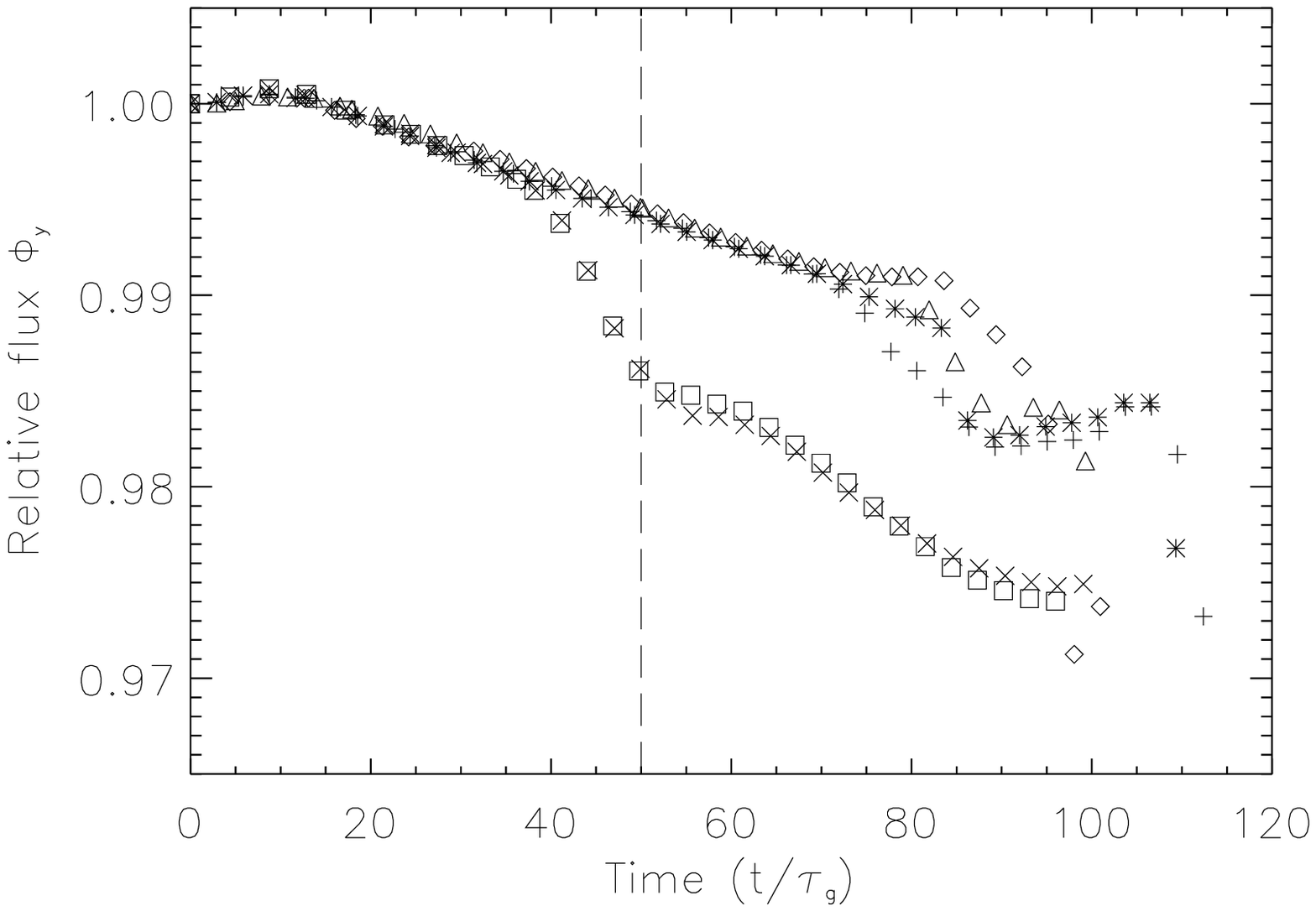}
\caption[]{Left: maximum axial magnetic field strength as function
of time (not height). The field strength for the same six models
as in Fig. \ref{fig2} are plotted with the field strength of the
reference model subtracted. Right: the magnetic flux $\Phi_y$
through the $xz$-plane as function of time for the same six
models as in Fig. \ref{fig2}. The vertical dashed line indicates
approximately the time when the ropes' centers pass the initial
lower boundary of the magnetized layer. } \label{fig7b}
\end{figure*}

Figure \ref{fig7b} (left) shows the decrease of the axial magnetic
field strength $B_z$ relative to the field of the reference model
as a function of time. The axial field of the reference model
decreases in a manner similarly to an adiabatically
expanding one-dimensional tube, see e.g.\ \cite{Dorch2002}. For
the models with non-zero $B_{\rm CZ}$, the deviation from the
reference model is substantial: for higher $\epsilon$ the field
decreases more slowly with time than the simple model. This is
because in the high $\epsilon$ models the flux ropes rise more
slowly and hence expand more slowly. In situations where the flux
ropes are halted, it is no longer the time scale of the rise that
dominates the decreasing field, but the diffusion time, which is
much longer. In case of low $\epsilon$ the rise is slightly faster
than the field-free reference case and then so are the ropes'
expansion and the decrease of the field strength implying negative
$B_z - B_{\rm ref}$ as seen in Fig.\ \ref{fig7b}.

In addition, the ropes that have an anti-parallel twisted
field component have lower axial field strengths than their
parallel-field counterparts when they begin to rise. The
explanation of this lies in what happens to the protective twist
that surrounds the flux ropes' cores (see the discussion below).

\section{Reconnecting the twist}

\begin{figure*}[!htb]
\centering
  \includegraphics[width=8.5cm]{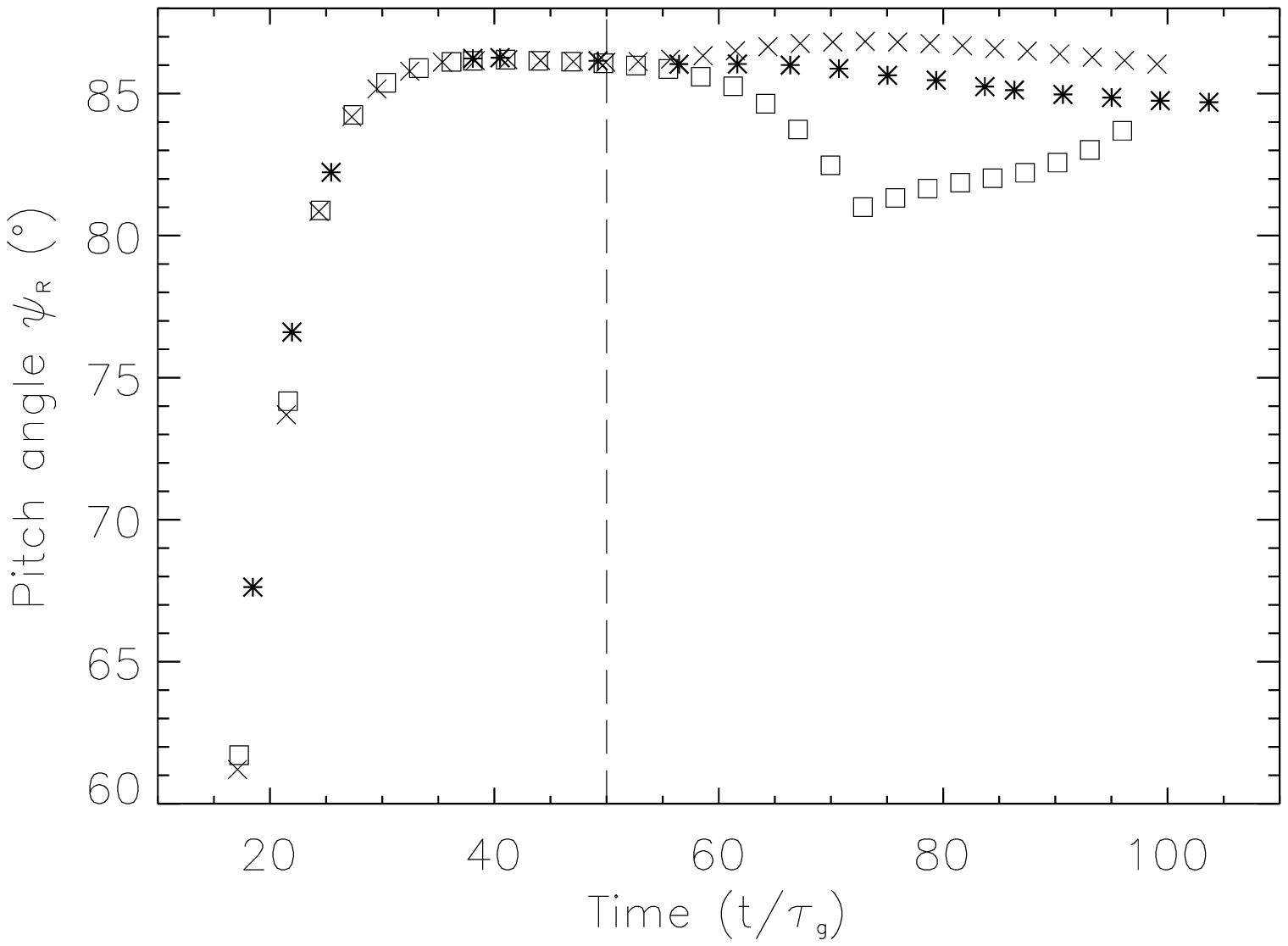}
  \includegraphics[width=8.5cm]{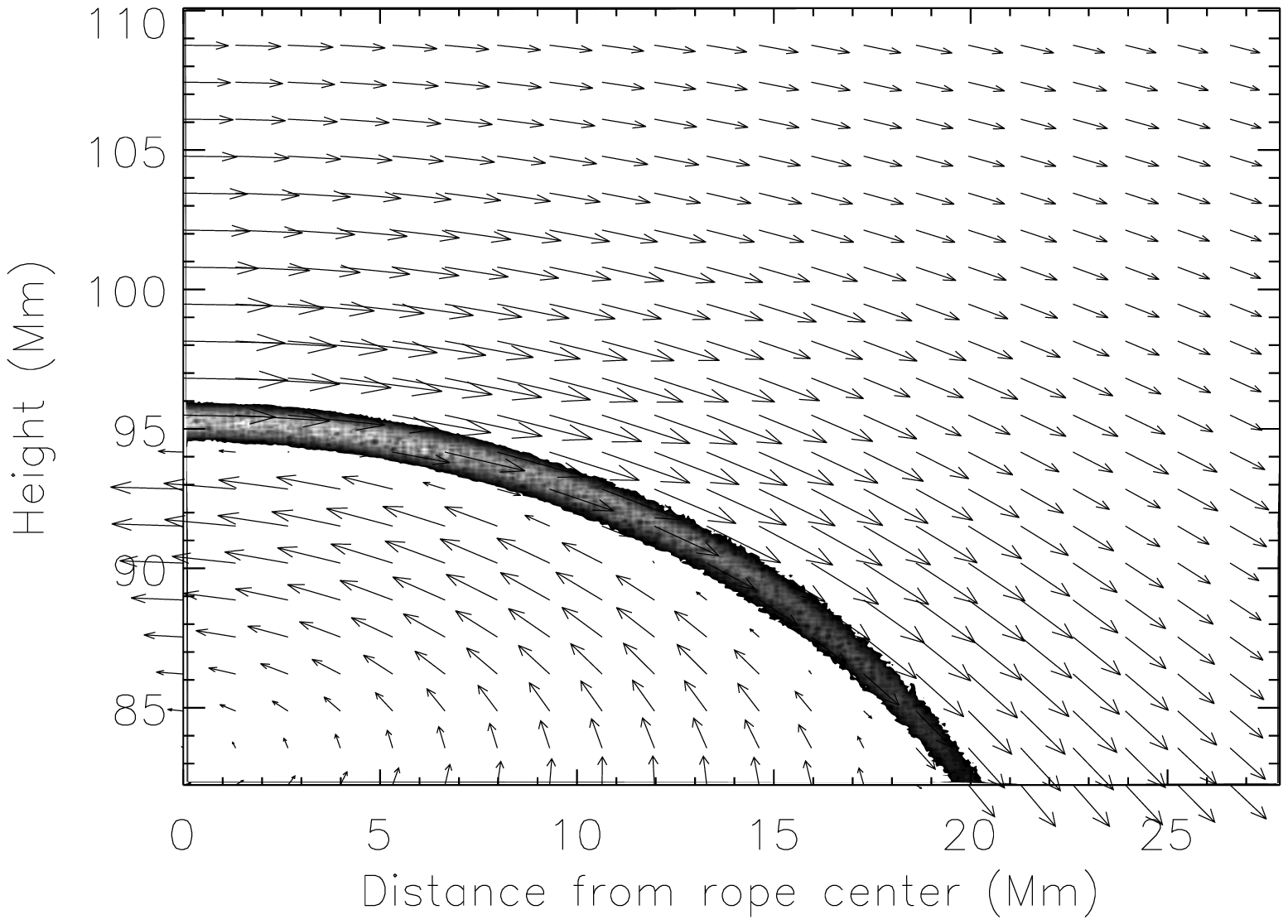}
\caption[]{Left: the maximum field line pitch angles are shown as
a function of time for the models 1A (squares) and 1B (crosses)
that correspond to $\epsilon = 0.044$ and oppositely directed
twisted field components. Also shown is the reference model 0
(stars). Right: subsection of the plane around the apex of the
rope in model 2A. Shown are the magnetic field (vectors) and
current density (grey scale). } \label{fig5}
\end{figure*}

Figure \ref{fig7b} (right) shows the flux through the $xz$-plane
(i.e.\ through the plane spanned by the flux rope's axis and
the vertical) as a function of time: at first one may find it
surprising that it seems that the low $\epsilon$ models loose flux
more rapidly than the ones with a stronger magnetized layer.
However, there are two effects that may cause a declining
$y$-flux: flux loss out of the computational domain through advection
and magnetic dissipation (diffusion). The
magnetic layer could in theory be expelled from the
computational domain via advection, but the average vertical
transport velocity on the top boundary remains negligible,
because of the closed upper boundary condition.

The second agent for reducing the $y$-flux is diffusion, but
reconnection between the twisted field of the flux rope and the
magnetic layer only changes the field lines' connectivity and does
not contribute to diffusion of the $y$-flux. In stead,
diffusion takes place at the bottom of the magnetic layer where
the vertical gradient of the field initially is very sharp. Hence,
natural physical diffusion causes the initial behavior of $\Phi_y$
to be identical for all the models up to around time equals 40--50
$\tau_g$: the flux decay is proportional to slow exponential decay
on a time scale of roughly 6000 $\tau_g$.

After the ropes begin to interact with the magnetic layer, the
results again depend on $\epsilon$: there is a noticeable
difference between the high and low $\epsilon$ models. As
mentioned, in the latter models $\Phi_y$ begin to decay more
rapidly, as the magnetic layer is mixed into the lower lying
atmosphere, which initially was field-free: this mixing introduces
further diffusion of the $y$-flux through new magnetic gradients
developing at small scales.

\begin{figure*}[!htb]
\centering
 \includegraphics[width=8.cm]{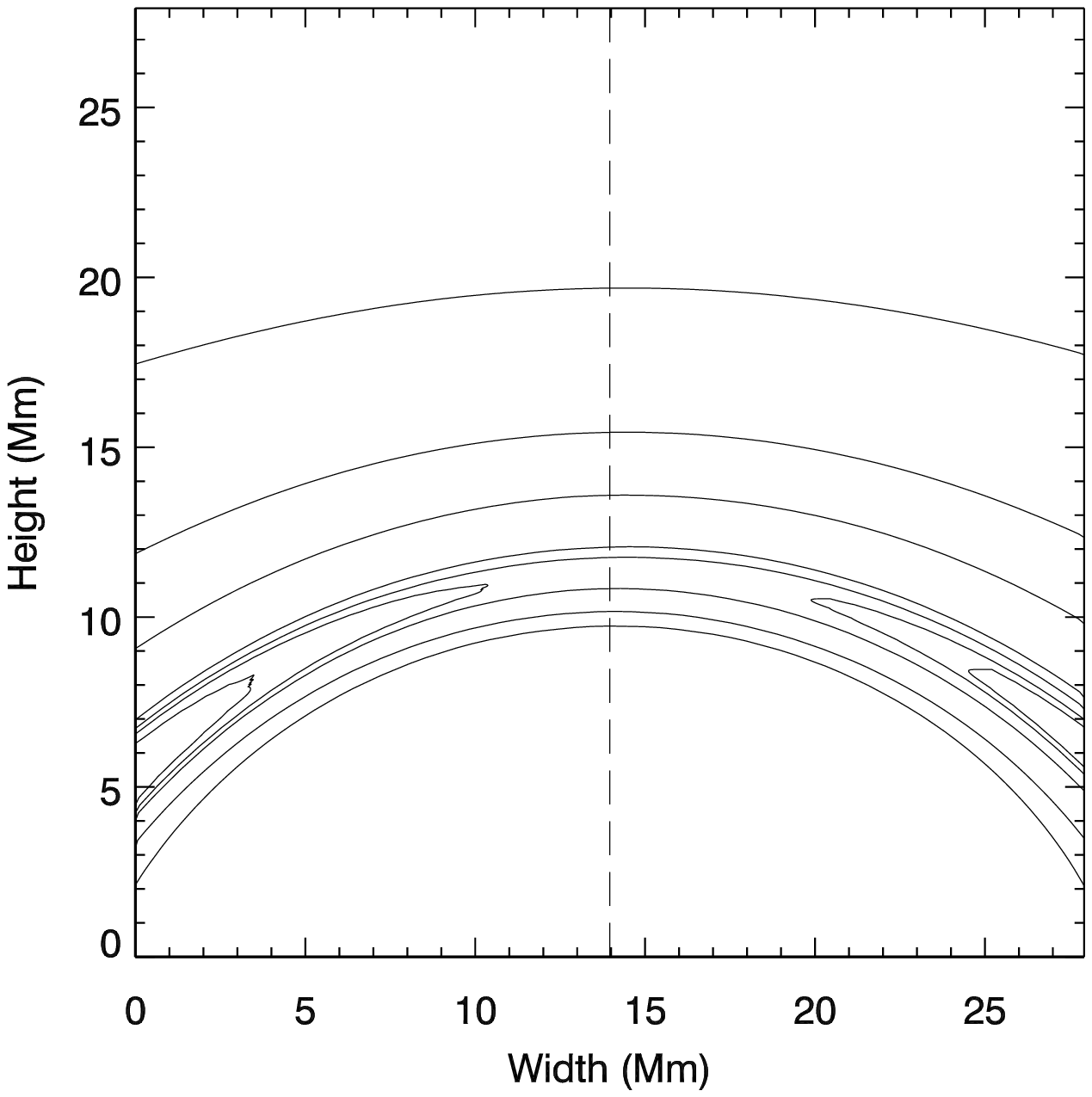}
 \includegraphics[width=8.cm]{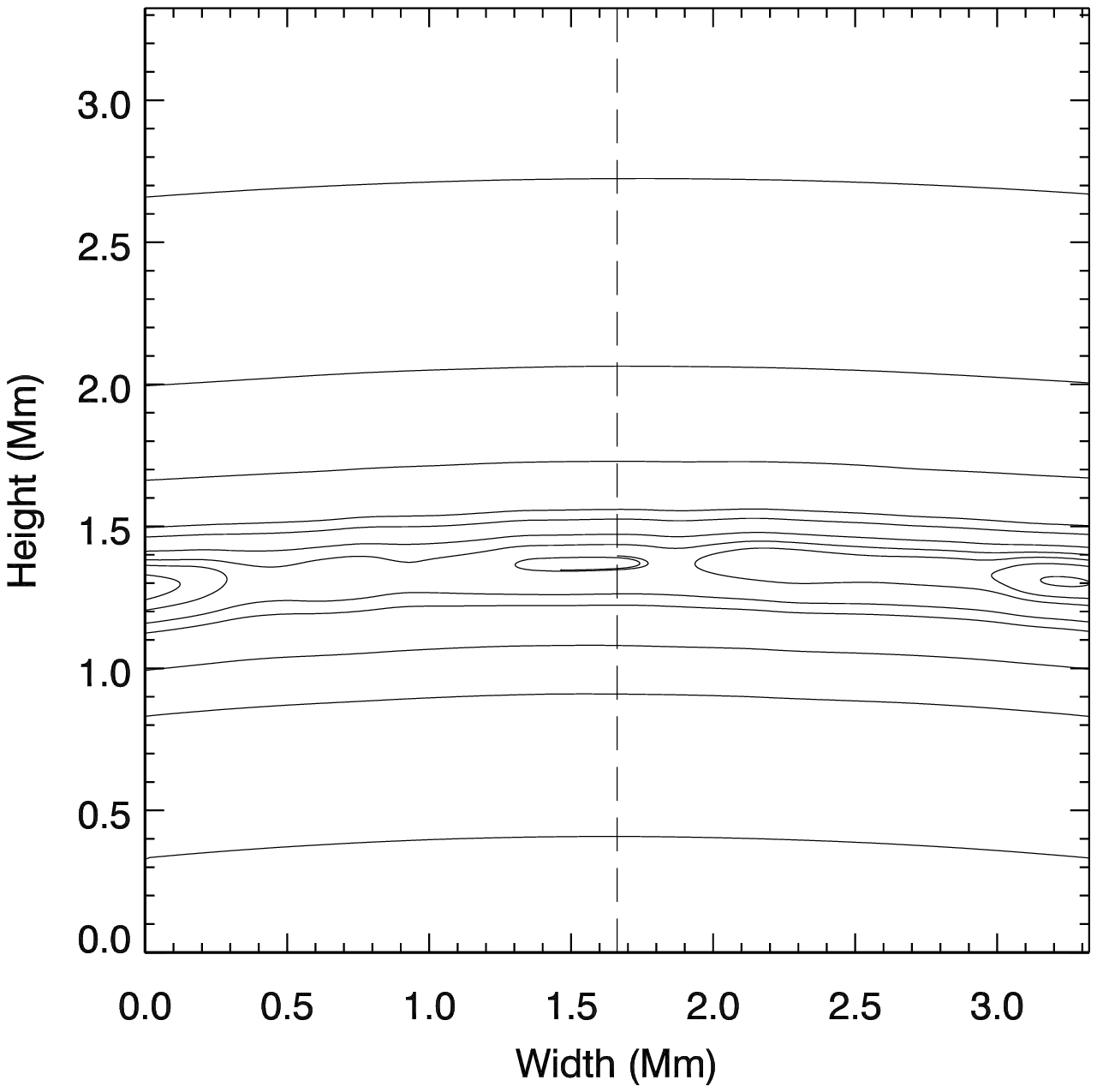}
\caption[]{Magnetic field lines near the front of the flux rope in
model 2A at time = 20. The upper most field lines belong
to the magnetic layer in the CZ and point to the right, whereas
the lower more curved field lines from the flux rope points
towards the left. Left: view of the rope's apex on a scale of
$\sim$ 25 Mm. Right: zoom at the apex where the field lines are
nearly horizontal.} \label{fig6}
\end{figure*}

Figure \ref{fig5} (left) shows the evolution of the field line
pitch angle $\Psi_{\rm max}$ for the two models with $\epsilon =
0.044$. These two models have oppositely directed twisted field
components compared to the direction of the envelope field. The
pitch clearly develops in a manner different from the field-free
reference model, and the evolution depends on the direction of the
twist: as expected the pitch of the rope in model 1B increases
beyond the reference model, as the parallel field lines of the
rope's twisted field is aligned up against the envelope field,
which has the same polarity just in front of the rope. In the
opposite case the pitch decreases when the rope in model 1A
encounters the magnetized layer: the core protecting twist is
gradually peeled way by field line reconnection, revealing the
less pitched field lines belonging to the `interior' of the rope.
It is this effect that causes the dependence on the sign of the
twisted rope component on e.g.\ the declining $y$-flux
$\Phi_y$ (Fig.\ \ref{fig5}) and the decrease of the axial field
strength $B_z$ (Fig.\ \ref{fig7b}, right).

Figure \ref{fig5} (right) shows a section around the apex of the
flux rope of model 2A: the magnetic field vectors change direction
across the current sheet that surrounds the upper part of the flux
rope. In Fig.\ \ref{fig6} (left), the field lines corresponding to
the vectors in Fig.\ \ref{fig5} are shown for a larger section
near the rope's apex. On this scale the behavior of the magnetic
field is very reminiscent of reconnecting field lines around a
magnetic null point, leading to an X-type topology: near the
central vertical line above the rope's center, the upper almost
stationary field meets an incoming oppositely directed field,
leading to reconnection across the null point which separates the
two field line systems. The reconnected highly bend field lines
move away almost horizontally. However, in Fig.\ \ref{fig6}
(right), the view is zoomed in on the current sheet by a factor of
10: the reconnecting region of the field clearly does not have a
simple X-type topology, but rather a more complicated situation
arises where closed magnetic field lines (plasmoids) are formed in
the center of the current sheet, e.g.\ \cite{Shibata+ea95}
and \cite{Archontis+ea06}.

The interaction of the flux rope with the overlying magnetic layer
not only changes the flux ropes' twist through field line
reconnection, but also alters the geometrical shape of the ropes:
the general tendency is that ropes with a twist directed parallel
to the CZ field become more roundish as they rise into and become
encapsulated in the magnetic layer, see e.g.\ the model 2B in
Fig.\ \ref{fig3}---ropes with anti-parallel twist loose some of
their protective transversal `shielding' and their axial fields
are much more steeply inclined at the ropes' upper boundaries
(e.g.\ model 2A in Fig.\ \ref{fig3}).

\section{Discussion and conclusion}
\label{discussion.sec}

It can be argued that two-dimensional models have several
restrictions compared to the full three-dimensional case: e.g.\
there is a tendency for enhanced flows due to the lack of the
extra degrees of freedom associated with three dimensions.
Therefore, one can consider the questions posed in this paper to
be addressed here only to a certain order.

On the one hand, the initial naive assumption that one could
have---that the effect of the magnetized layer would be to destroy
a flux rope with a twist anti-parallel to the polarity of the
layer---is wrong. In fact, the opposite is true: even though
reconnection within the magnetic layer reduces the twist of the
flux ropes, it is easier for them to penetrate the poloidal layer,
because the reconnection of field lines allows the ropes to
``carve'' their way through the layer. This is provided, however,
that the flux ropes contain enough transversal flux.

On the other hand, the simple estimate of $\chi_{\rm c} \approx 5$
in Eq.\ \ref{chi_c.eq} seems to hold in the simulations: in the
models with $\chi < \chi_{\rm c}$, the ropes' buoyancy are
strongly damped. However, while ropes with $\chi > \chi_{\rm c}$
are allowed to rise, they may still be halted in the CZ, but the
more correct critical limit will have to be a function also of
twist, i.e.\ $\chi_{\rm c} = \chi(\epsilon)$.

It has been shown here that the flux ropes' dynamics depend on
both the strength and sign of the poloidal field: for a relatively
strong poloidal field the rise is slower and therefore the flux
ropes reach a lower height in the same amount of time. In certain
cases the rise can be completely halted. It turns out that
anti-parallel twisted ropes reach higher and faster. When it comes
to the flux ropes' topology, the geometrical shape of ropes also
depends on the strength and sign of the poloidal field: the apex
of anti-parallel twisted ropes are flatter and have steeper
magnetic gradients in their axial field. Ideally this could be
observed as a seemingly faster passage through horizontal layers
(emergence) of the axial part of the flux rope, relative to cases
with more moderate vertical gradients in the axial field: e.g.\
\cite{Archontis+ea04} have shown that in fact a steep axial
gradient is required by flux emergence (their Eq. 10). As to the
flux ropes' twisted field components, reconnection reduces the
twist of anti-parallel ropes. If ropes on either sides of the
equator are oppositely twisted, this could then in principle be
falsified observationally by comparing observations of poloidal
field strength in emerging northern and southern active regions.

The most fundamental problems remaining are those of the origin of
the twist, and the question of how it arises. One may speculate
that twisted field lines could be generated in large-scale flux
bundles located near the bottom of the convection zone connecting
across the solar equator: such flux bundles would experience a
rotating motion since their lower parts are located in a region
rotating slower than their uppermost parts. This rotation would
twist the magnetic field lines and this twist could be transmitted
to the parts of the flux bundle at slightly higher latitudes,
thereby possibly giving rise to a twisted toroidal flux system.
The sign of a twisted component generated in this geometric manner
would always be pointing in the same direction (northwards in the
Sun). Hence, in consecutive 11-year half-cycles of the solar
dynamo, the twist would alternate between being anti-parallel and
parallel to the poloidal field (from which the corresponding
toroidal field was generated in the respective cycle). Such an
alternation would result a 22-year cycle in the amount of twist in
the emerging parts of the toroidal flux ropes, e.g.\ in the
helicity of bipolar magnetic regions.

Fully three-dimensional MHD simulations extending the present study
are under way, using the numerical scheme of \cite{Archontis+ea04}.

\begin{acknowledgements}
Access to computational resources granted by the Danish Center for
Scientific Computing is gratefully acknowledged.
\end{acknowledgements}

\end{document}